# Causation of Late Quaternary Rapid-increase Radiocarbon Anomalies

G. Robert Brakenridge

Institute of Arctic and Alpine Research (INSTAAR), University of Colorado, 4001 Discovery Drive., Boulder CO 80303

Phone (603)2520659, Email Robert.Brakenridge@Colorado.edu

## Abstract

Brief (<100 y) rapid-increase anomalies in the Earth's atmospheric $^{14}C$ production have previously been attributed to either γ photon radiation from supernovae or to cosmic ray particle radiation from exceptionally large solar flares. Analysis of distances and ages of nearby supernovae (SNe) remnant surveys, the probable γ emissions, the predicted Earth-incident radiation, and the terrestrial $^{14}C$ record indicates that SNe causation may be the case. SNe include Type Ia white dwarf explosions, Type Ib, c, and II core collapse events, and some types of γ burst objects. All generate significant pulses of atmospheric $^{14}C$ depending on their distances. Surveys of SNe remnants offer a nearly complete accounting for the past 50000 y. There are 18 events ≤ 1.4 kpc distance, and brief $^{14}C$ anomalies of appropriate sizes occurred for each of the closest events (BP is calendar years before 1950 CE): Vela, +22‰ $\Delta^{14}C$ at 12760 BP; S165, +20‰ at 7431 BP; Vela Jr., +13‰ at 2765 BP; HB9, +9‰ at 5372 BP; Boomerang, 11‰ at 10255 BP; and Cygnus Loop at 14722 BP. Although uncertainties remain large, the agreements of prediction to observation support a causal connection.

**Introduction**

Several very brief (< 100 y), rapid-increase (within 2 y) positive anomalies in the Earth's late Quaternary atmospheric $^{14}$C concentration have been attributed to nearby supernovae (SNe) (Damon et al. 1995); to γ-ray burst (GRB) objects (Pavlov et al. 2013); or to exceptionally large solar flares (Miyake et al. 2014; Usoskin et al. 2013). Although the sizes of the anomalies require exceptionally large solar radiation events, there is also uncertainty whether γ emission from SNe could have reached sufficient intensities (Miyake et al. 2012). SNe causation has also sometimes been excluded due to lack of specific candidates (Usoskin et al. 2013). However, the radio and high energy surveys of supernovae remnants (SNRs) now include some newly-discovered objects (Kothes 2003) and an examination of the suite of relatively nearby SNes as compared with the known $^{14}$C anomalies has not before been accomplished.

SNe are diverse, may include many GRBs, and total energies vary between $10^{49}$-$10^{54}$ ergs. Such diversity is likely the case also for their total γ emissions. There are presently at least 383 supernova remnants (SNRs) in our galaxy (Safi-Harb et al. 2012) and most are younger than $5 \times 10^4$ y, thereby matching in time the interval for which records of $^{14}$C anomalies are available. There are thus many possible SNe candidates for $^{14}$C perturbations and their ages and distances, while not known precisely, are at least increasingly well-constrained. This paper explores which of these SNe may be of appropriate energies, distances, and ages to have caused the recorded $^{14}$C changes.

*SNe* as used here include: Type Ia white dwarf explosions (Churazov et al. 2015), Type Ib, Ic, and II massive star core collapse events (Podsiadlowski 2013), core collapse hypernovae (Pian et al. 2006), superluminous SNe (Moriya et al. 2018), many or most long GRBs (Cano 2014; Cano et al. 2017; Gehrels and Mészáros 2012), and subluminous long GRBs (Nakar and Sari 2012). Prompt isotropic relativistic shock breakout γ emissions of $10^{48}$ and $10^{48-50}$ ergs from "standard" long GRBs (including those with beamed emissions) and also from subluminous GRBs is predicted by theory and agrees with the limited observational data (Nakar and Sari 2012). Type II SNe are theorized and modeled to produce high energy photon emissions through prompt breakout emissions (Colgate 1975; Klein and Chevalier 1978) and γ emission may also be sustained over several years by other mechanisms. Observations in the γ domain of the nearby SN 1987 in the large Magellanic Cloud tested some of the relevant theory and observed such γ (Matz et al. 1988). Hard photon emissions are also being observed in association with the several types of SNe in other galaxies by the orbital high energy observatories; the γ spectrum and energy for a Type Ia event was recently obtained (Churazov et al., 2015). These data, theory, and modeling place constraints on the Earth-incident radiation to be expected from the objects recorded by the galactic SNRs. However, they also may broaden these constraints, as the observational variety of SNe is becoming more complex.

SNe hard photon effects on solar system planetary bodies are mediated by any atmosphere, which shield the surfaces, but absorb, scatter, and re-emit the radiation. For nearby SNe, the associated γ radiation is predicted to affect Earth's atmosphere in measurable ways (Gehrels et al. 2003; Ruderman 1974): including through production of cosmogenic isotopes (Menjo et al. 2005; Pavlov et al. 2013). In this regard, steady levels of radioactive $^{14}C$ are maintained in Earth's upper atmosphere by effects on N from incoming galactic and solar cosmic ray particles. However, 10-40 Mv γ photons from SNe can also produce $^{14}C$, $^{10}Be$, and $^{36}Cl$, ionize N by photonuclear reactions, and initiate neutron cascades (Damon et al. 1995; Lingenfelter and Ramaty 1970;

Miyake et al. 2012). Thermalized neutron yields from γ photons reach a maximum at about 23 MeV (from absorption around the giant dipole resonance for N and O nuclei) (Pavlov et al. 2013). $O^3$, an important greenhouse gas and solar UV shield, may be depleted also by the ionizing radiation, and catalytic reactions producing $NO_x$ species initiated (Gehrels et al. 2003; Ruderman 1974; Thomas et al. 2005). For planetary bodies and comets unprotected by an atmosphere, intense γ radiation may induce rock surface melting or ice volatilization: nearby GRBs have sometimes been considered as an origin for solar system chondrules (Scalo and Wheeler 2002); possible effects on icy planetary surfaces remain unexplored. At present, even extragalactic GRBs are producing small but measureable effects on the Earth's ionosphere (Fishman and Inan 1988). The Earth's exposure to SN-originated γ radiation must therefore be included as a component of its overall environmental history, and the impact of these rare but extreme radiation events may extend to other planetary bodies in the solar system.

Here are examined possible terrestrial isotope ($^{14}C$) records from known galactic SNe. Note that, at ages $> 5 \times 10^4$ y, expanding SNRs blend into the interstellar medium, and the opportunity to examine specific events at constrained times and distances is less robust. Analysis here is confined to the younger time period and to one cosmogenic isotope: $^{14}C$. Also, other papers explore the possible effects on $^{14}C$ of cosmic radiation (particles) from SNe and SNRs (Firestone 2014), but these travel at less than the speed of light and arrive at Earth $10^2$ to $10^3$ y after the γ photons. This paper is restricted to the predicted effects from the photons.

**Locating Nearby Supernovae**

Three comprehensive radio and high energy catalogs of SNRs (Green 2014; Pavlović et al. 2014; Safi-Harb et al. 2012) were interrogated to identify objects $< 5 \times 10^4$ y in age and <1.5 kpc in distance, providing 18 objects (Table 1). Although some ages for radio SNRs are based on surface brightness/remnant diameter (Σ–D) and D-age relations, these are calibrated using measured radial expansion velocities and other methods and also their precision is known to be low (Pavlović et al. 2014). Thus, these were used only when no other observation-based estimates were available. Where uncertainty values are published, they are included in Table 1; otherwise "~" is indicated and an uncertainty of ± 25% (standard error) is assumed; all uncertainties are carried through to the energy and $^{14}C$ production calculations. There are also approximately 20 poorly-constrained SNR objects in the high energy compilation for which no distances or ages are available. These were excluded for the purpose of this paper and because it is unlikely that many of these are close, relatively young, and of importance to the present analysis.

| Catalog Number | Distance (kpc) | Total γ (ergs/cm$^2$) | SN Age (BP) | Predicted $^{14}C$ Production (a/cm$^2$/s) | Measured Δ$^{14}C$ Anomaly and Age Range |
|---|---|---|---|---|---|
| G263.9-03.3 | .25 ± .03 | 4.3-6.9 x 10$^6$ | 14500 ± 1500 | 17.6-28.5 | +40‰, 12760-12630 BP |
| G330.0+15.0 | .32 ± .17 | 1.4-15 x 10$^6$ | 23000 ± 8000 | 5.7-61.2 | +21‰, 22500-22360 BP |
| G114.3+00.3 | .70 ± .35 | 3.0-27 x 10$^5$ | ~7700 | 1.2-11.2 | +20‰, 7431-7421 BP |
| G266.2-1.2 | .70 ± .25 | 3.7-16 x 10$^5$ | 3800 ± 1400 | 1.5-6.8 | +13‰, 2765-2749 BP |
| G074.0-08.5 | .74 ± .03 | 5.6-6.6 x 10$^5$ | 15000 ± 5000 | 2.3-2.7 | 14722-14712 BP |
| G160.9+02.6 | .80 ± .40 | 2.3-21 x 10$^5$ | 5500 ± 1500 | 1.0-8.6 | +9‰, 5372-5362 BP |
| G106.3+02.7 | ~.80 | 3.3-9.3 x 10$^5$ | ~10000 | 1.4-3.8 | +12‰, 10255- 10220 BP |
| G040.5+00.5 | ~1.00 | 2.1-5.9 x 10$^5$ | ~20000 | 0.9-2.4 | |
| G190.9-2.2 | ~1.00 | 2.1-5.9 x 10$^5$ | ~1550 | 0.9-2.4 | +15‰, 1176-1166 BP |
| G152.4-2.1 | ~1.00 | 2.1-5.9 x 10$^5$ | ~6900 | 0.9-2.4 | |
| G107.5-1.5 | ~1.10 | 1.8-4.9 x 10$^5$ | 4500 ± 1500 | 0.7-2.0 | +20‰, 4880-4820 BP |
| G127.1+0.5 | ~1.15 | 1.6-4.5 x 10$^5$ | ~25000 | 0.7-1.9 | +46‰, 26200-25520 |
| G205.5+0.5 | 1.20 ± .40 | 1.3-5.2 x 10$^5$ | 90000 ± 60000 | 0.5-2.2 | |
| G347.3-00.5 | 1.30 ± .40 | 1.2-4.1 x 10$^5$ | 1840 ± 260 | 0.5-1.7 | +9‰, 957-947 BP |
| G180.0-1.7 | 1.30 + .22, -.16 | 1.4-2.6 x 10$^5$ | 30000 ± 4000 | 0.6-1.1 | |
| G260.4-3.4 | 1.30 ± .30 | 1.3-3.3 x 10$^5$ | 1990 ± 150 | 0.5-1.4 | |
| G119.5+10.2 | 1.40 ± .30 | 1.2-2.8 x 10$^5$ | ~13000 | 0.5-1.1 | |
| G327.6+14.5* | 1.56 | 1.4-2.4 x 10$^5$ | 994 (SN 1006) | 0.4-1.0 | +8‰, 942-933 BP |

(Table Caption at Manuscript End)

A 4 x $10^{49}$ ergs estimate is used in this paper for typical total (isotropic) SN γ energy as emitted over each complete event and with a nominal duration of 1 y. It is based on theoretical, observational, and modeling results for both galactic and extragalactic SNe, as further described below. However, individual event intensities, duration, and γ spectrum may vary widely. This value is thus used to identify the most important nearby events in the table, but actual excess production of $^{14}$C by each may have been much larger or smaller. These uncertainties may be reduced in the future by consideration of the individual candidate SNR characteristics, such as calculated total explosion energies and progenitor star masses.

**Determining SNe γ Emission Energies**

Earth's SNe hard photon radiation history is a function of the emission energies of SNe events and their distances (the galactic interstellar medium is relatively transparent to γ photons). The uncertainties concerning intrinsic emission energies require further discussion, as their possible range is critical to understanding the Earth's late Quaternary exposure. Decades prior to any observations, SNe theory for core collapse events predicted prompt X- and γ radiation. Peak luminosities of 1.9 X $10^{45}$ ergs s$^{-1}$ were calculated, and the total hard γ energy from SNe was estimated to vary between $10^{47}$ ergs and $10^{50}$ ergs, radiated over a period of months (Colgate 1975; Klein and Chevalier 1978). Type II core collapse SNe have sometimes been used as "standard candles", but comparisons with observational data for extragalactic objects show that their total explosion energies vary from 0.5 to 4.0 x $10^{51}$ ergs (Kasen and Woosley 2009). Perhaps only .01 of such energy is emitted as γ (Miyake et al. 2012). Varying SNe γ emissions depend on the mass and type of the progenitor star, metallicity, rotation velocity, and whether a binary system is involved (Kann et al. 2018).

During an SN, an initial shock breakout may produce either beamed or isotropic γ emission. Observations with γ and X-ray observatories and optical telescopes demonstrate that many long (10-300 s) GRBs are a special class of extra-galactic, supermassive star SN with beamed γ emission reaching isotropic-equivalent energies of $10^{53}$ ergs (Gehrels and Mészáros 2012). At least one late Quaternary galactic SNR exhibits characteristics compatible with origin as a GRB (Lopez et al. 2013). Also, XRFs (e.g., SN 2006j) produce prompt X-ray flashes. Such objects are half as luminous as some GRB-associated optical SNe; they attain total energies smaller than GRBs but greater than typical SNe, and may be isotropic radiators of γ and X-rays also (Pian et al. 2006). Optically, "superluminous supernovae" are another class of observed objects and exhibit peak brightnesses approximately 10-100 times that of more common SNe (Moriya et al. 2018). Their γ emissions may be much larger as well. Finally, for Type Ia (binary white dwarf) supernovae, observation of an extragalactic example indicates γ luminosities of $11 \pm 1 \times 10^{41}$ erg s$^{-1}$ on day 73 and $6.5 \pm 0.6 \times 10^{41}$ erg s$^{-1}$ on day 96 (Churazov et al. 2015). A year of such emission would provide a total γ reaching near ~$1 \times 10^{49}$ erg (depending on the size of the earliest emission).

In regard to core collapse SNe, Many GRBs and superluminous SN may be hypernovae producing black holes (Podsiadlowski 2013): unusually energetic core collapse SNe with supermassive star progenitors. Prompt emission in γ may be from successful or failed shock breakout (Nakar and Sari 2010), some days prior to initiation of the optical event. Then, as the explosion evolves, γ radiation again emerges, and is sustained over a period of several years and dependent on characteristics of the expanding shell (Matz et al. 1988). The spectrum for SN 1987A, a well-observed and relatively nearby core collapse SN in the Large Magellanic Cloud, provides an example of that sustained emission. Hard photon emission was observed between 0.02 and 2 MeV over 500 days; the measured total γ energy was $10^{46}$ ergs and total SN energy was 1.4 ±.6 x $10^{51}$ erg (Chevalier 1992; Pinto and Woosley 1988). However, the progenitor for 1987A was a blue supergiant with an initial mass of about 20 M☉ instead of the more typical red supergiant, and the SN was fainter than typical Type II SNe at maximum by an order of magnitude (Chevalier 1992). Thus, modeling of possible terrestrial γ effects (Gehrels et al. 2003) of "typical" SNe used a higher ($10^{47}$ erg) γ total, a spectral distribution binned into 66 logarithmic intervals 0.001-10 MeV, and assumed a red supergiant progenitor of 15 M☉. For comparison, and as regards known late Quaternary supernovae remnants, the Vela SN's progenitor was 30 M☉ (Sushch and Hnatyk 2014). In the modeling, the Type II SN γ luminosity peaks at 340 d and is within a factor of 10 of the peak for 500 days.

This paper adopts 4 x $10^{49}$ ergs as a reasonable average total γ (including prompt relativistic shock breakouts, any intercepted jetted emission, and sustained emission) for comparing prehistoric SNe events to the terrestrial $^{14}$C record. Total γ emission from, in particular, supermassive star core collapse events may reach such a total and including very quickly in isotropic breakouts. For example, the low-luminosity long GRBs exhibit an overall isotropic equivalent radiated γ energy of ≲ $10^{49}$ erg. Although rarely observed in other galaxies because of their low luminosity, they are more numerous than regular long GRBs in terms of rate per unit volume (Nakar and Sari 2012). To consider all of the SNRs listed in Table 1 as derived from similar events is probably unrealistic. However, as knowledge and theory of optical and emitting SNe in other galaxies expands, it may soon be possible to apply observational criteria (Lopez et al. 2013) to SNRs to better determine their individual parameters and their associated radiation-emitting histories.

**Comparing Predicted Terrestrial γ Incidence to the $^{14}$C Record**

Earth-incident SNe radiation varies with $1/d^2$, and distance (d) estimates include large uncertainties for many objects (Table 1). They are based on a variety of observational methods: proper motions, shock and radial velocities, HI absorption and polarization, kinematic spectral line observations, and association with star fields measured via parallax. Accuracies vary; for kinematic distances, the uncertainties may be <30%; for distances from X-ray fitting, they may be >50% (Zhu and Tian 2013). Where uncertainties are not provided for particular objects, a ± 25% value is assumed and represents an approximate average of the published uncertainties (Table 1).

Available age estimates are partially dependent on measured or estimated distances. Those in Table 1 include ages from empirical SNR radio surface brightness/remnant diameter Σ–D and D–age relations, and also from measured radial expansion velocities; the latter may be more accurate. Where uncertainty values are published, they are included in the Table. The most recent observational findings for the SNe are used in each case.

To understand needed SNe γ requirements for causality, the relation between observed $^{14}$C changes, commonly expressed as Δ $^{14}$C (Stuiver and Polach 1977), and $^{14}$C production in the upper atmosphere must be modeled. Carbon cycle modeling, including the pathways of atmospheric $^{14}$C and its incorporation into terrestrial records, is increasingly comprehensive (Kanu et al. 2016). However, the purpose of this paper is the identification of important candidate SNe that may have affected global production. Therefore, the published results of relatively simple 4- or 5-box models are used to compare the SN-predicted increases in $^{14}$C production with observed biosphere Δ $^{14}$C changes (Table 1 and text below). Note that, unlike particle cosmic radiation, Earth-incident γ is not affected by the geomagnetic field. However, complex atmospheric changes that may be initiated by γ radiation could also themselves affect the resulting $^{14}$C record (Pavlov et al. 2013).

## The Historical SN 1006 Example

The possibility of SNe affecting Earth's $^{14}$C production has been investigated intermittently for over 4 decades, even though many of the relatively nearby objects in Table 1 have not been considered. For example, in tree ring records, a 9.5‰ Δ$^{14}$C rapid-increase tree ring anomaly commences at 942 BP (BP = y before 1950 CE). This is 2 y after the historic SN 1006 (Damon et al. 1995; Lingenfelter and Ramaty 1970). For SN 1006, a distance of 1.3 kpc was used to calculate Earth-incident γ, and an intrinsic energy of 1 x 10$^{49}$ ergs in γ >10 Mev. This produces 1.4 x 10$^4$ ergs cm$^{-2}$ at Earth and yields approximately 10$^3$ neutrons per erg (Lingenfelter and Ramaty 1970); 65% was assumed to be thermalized and available to produce $^{14}$C by $^{14}$N(n,p)$^{14}$C. Thus, .9 x 10$^7$ thermal neutrons generate the SN-related $^{14}$C (Damon et al. 1995). If this arrives in one year, the $^{14}$C production is .3 a/cm$^2$/s, as compared to the annual steady-state production by cosmic rays of 1.64 a/cm$^2$/s (Kovaltsov et al. 2013). On the other hand, the observed tree ring-recorded anomaly, which decays over 9 y, was fit via carbon cycle box modeling to a one year only, 2.5x increased $^{14}$C production rate (Damon et al. 1995). A subsequent tree ring search for the same $^{14}$C anomaly at another geographic location was successful; a +5‰ Δ$^{14}$C increase was measured (Menjo et al. 2005). For a causal SNe connection, however, it appears that the SN energy must be larger than that assumed.

The distance of the SN 1006 SNR (G327.6+14.5, pulsar PKS 1459–41) has since been revised to 1.6 kpc (Jiang and Zhao 2007). Also, a much lower, 20–55 a/erg mean yield is used by (Pavlov et al. 2013) for the gamma-ray flux entering the atmosphere from a hypothetical GRB with typical spectral parameters. Other recent studies use a production rates of 130 a/erg, (Matz et al. 1988; Usoskin et al. 2013) for γ-related production. This value is used in Table 1 for all events, again for comparison purposes. For SN 1006 and the associated SNR, and using a $4 \times 10^{49}$ γ energy, the 130 a/erg production rate, and the revised distance, the predicted results still produce a relatively small anomaly, but perhaps one that remains compatible with the $^{14}C$ observation (given the major uncertainties for γ emission sizes, cross sections, and $^{14}C$ production function). Thus, the calculated extra production of .4 to 1.1 a/cm$^2$/s (Table 1) is comparable to the modeled need for 2.1 a/cm$^2$/s (for the smaller 5‰ $\Delta^{14}C$ increase measured at the second site).

The tree ring-based $^{14}C$ concentrations reflect tropospheric conditions during wood formation, and radiocarbon is produced mainly in the stratosphere, so that some time lag is expected; such lags are variously accommodated by different box models (Pavlov et al. 2013). Also, SN 1006 is considered to be a Type 1a white dwarf event, which may have re-brightened (Jiang and Zhao 2007); not all of the γ may have been produced in one year. Thus, the observed anomalies at 942 BP at two sites (Damon et al. 1995; Menjo et al. 2005) may possibly record SN 1006 SN. Alternatively, an extreme solar flare could be invoked. Additional tree ring studies are needed in any case to further validate the event as global in geographic extent. Now are examined much closer events, most of which are in the prehistoric record.

**Comparison of Nearby SNe And Late Quaternary $^{14}$C Anomalies**

Prior examinations of the capability of SNe to affect Earth's atmosphere have mainly focused on historical SNe (Damon et al. 1995; Miyake et al. 2013; Miyake et al. 2016; Miyake et al. 2012); none of these were exceptionally close to the Earth. However, there are 18 SNe with distance estimates ≤ 1.4 kpc and ages less than 5 x 10$^4$ BP (Table 1). One is as close as ~ .25 kpc. Are the predicted $^{14}$C effects of these relatively nearby SNe compatible with the observed $^{14}$C record?

IntCal13 (Reimer 2013) and additional published tree ring based records are used here for part of this analysis. Dendrochronologically-dated tree rings provide the assayed carbon for the younger, <12,500 BP part of IntCal13. Other materials sample $^{14}$C in the upper mixed ocean (marine corals and foraminifera), soil water (speleothems), or lake biota (Southon et al. 2012). The complete IntCal13 temporal coverage is 50,000 BP to present, with much loss of temporal resolution and resulting attenuation of any brief pulses of atmospheric $^{14}$C in the older part of coverage. Thus, the temporal sampling is 20 y from 26000 to 15020 BP, 10 y from 15000 to 12500 BP, and 5 y from 12495 to 0 BP (Reimer 2013). IntCal13 may not detect short term variations lasting less than 500 y earlier than 15000 BP at all, unless they are exceptionally large. With these constraints in mind, the closest late Quaternary SNe are now compared to the $^{14}$C record and by use of more temporally detailed assays when available. Note that the ages provided for each event are those of the initiation of the 14C anomaly and are in in calendar years BP.

*G263.9-03.3, Vela, 12760 BP.* The nearest of late Quaternary SNe, the Vela core collapse SN, occurred (within uncertainty) at the same time as the largest rapid-increase global $^{14}$C anomaly (Table 1, Figure 1a). Its distance of .25 ± .03 kpc is from Ca II absorption line spectra (Cha et al. 1999) and there is an independent distance estimate of .29 ± .02 kpc from parallax measurements on the associated pulsar (Dodson et al. 2003). The age is estimated as 14500 ± 1500 (Cha et al. 1999); the pulsar characteristic age is 11400 y. A +22‰ $\Delta^{14}$C increase occurs within 60 y starting at 12760 BP in floating chronology tree ring records with close-interval sampling; +40‰ within 130 y (Hua and al 2009; Kromer et al. 2004). The IntCal13 curve provides instead a +25.3 ‰ $\Delta^{14}$C from 12745-12640 BP, and increasing another 5.4 ‰ to 12515 BP (Figure 1a). These latter results are based on decadal samples, dampen any short-lived peaks, and do not capture individual years. However, carbon cycle considerations and known lags in cross-hemisphere atmospheric mixing also suggest that several years would be required for full incorporation of the pulse into the global atmosphere and thence into $^{14}$C-recording materials such as tree rings.

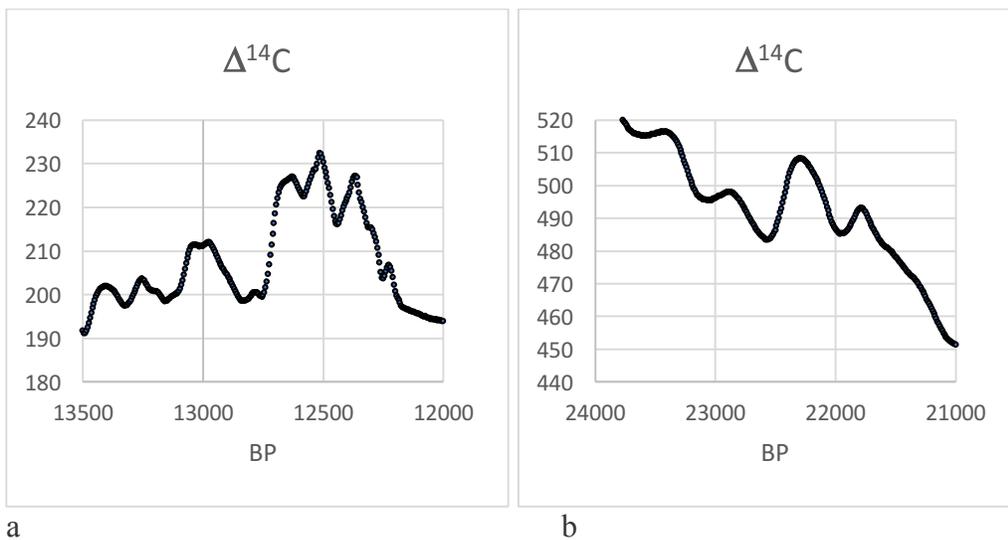

a                                              b

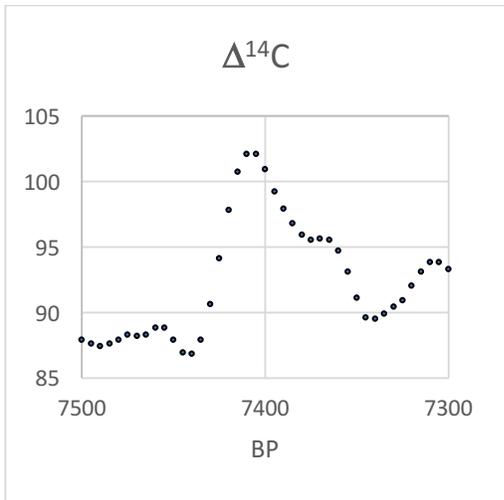

c

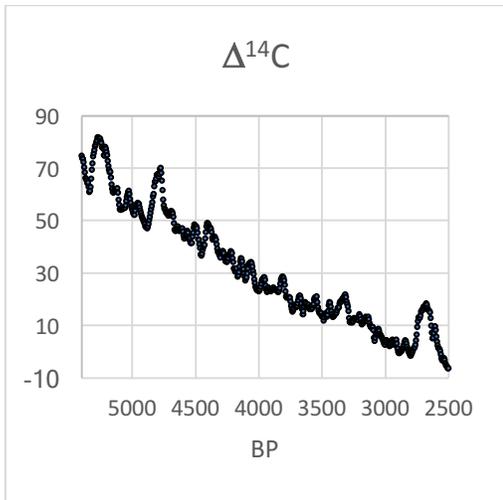

d

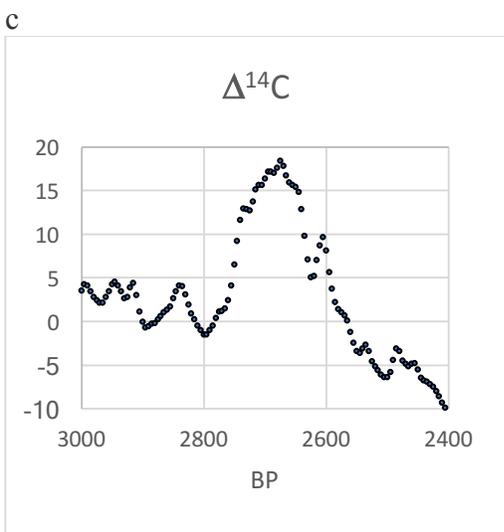

e

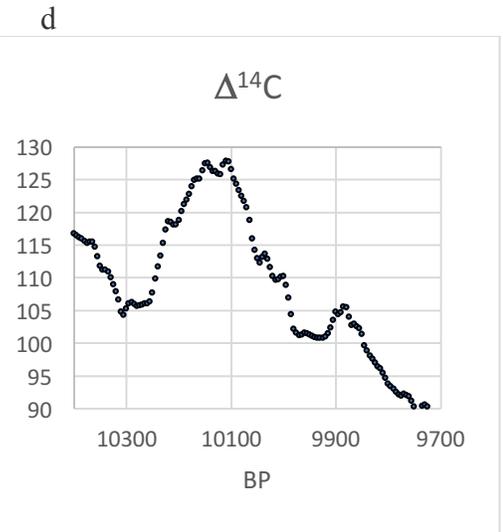

f

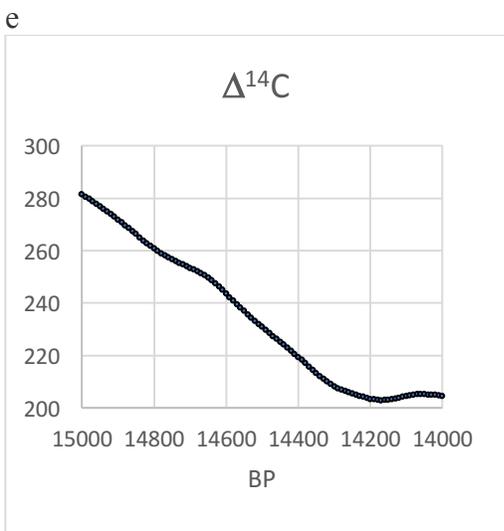

g

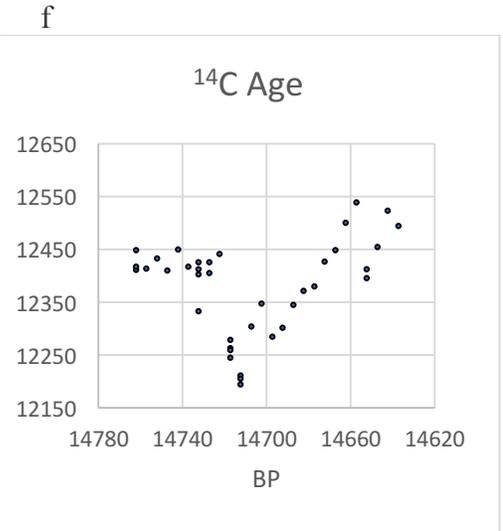

h

(Figure Caption at Manuscript End)

The timing of the $\Delta^{14}C$ increase is apparently synchronous with abrupt terrestrial climatic changes at the onset of the Younger Dryas Stadial: an interval of sharply cooler temperatures, especially at temperate to high northern latitudes (Hughen et al. 2000). This agrees with predictive modeling of climate cooling possibly caused by such a radiation event (Thomas et al. 2005): increased $NO_x$-induced atmospheric opacity and reduction of $O_3$, which is an important greenhouse gas, both favor cooler temperatures. Causation of the steep rise of $^{14}C$ is, however, controversial (Olivier et al. 2001). Climate-induced changes in oceanic circulation may have been involved: the high $^{14}C$ has been interpreted as the result of a reduced surface water exchange with the older, deep-ocean reservoir. The very rapid and large increase, observed in both tree rings and varved marine sediments (Hughen et al. 2000), is, however, difficult to model with only changes in oceanic circulation (Olivier et al. 2001), as follows:

*"The second feature in the Cariaco basin $\Delta^{14}C$ record not replicated by our model is the rapidity of the $\Delta^4C$ increase at the onset of the Younger Dryas…If the rapid $\Delta^{14}C$ increase at the onset of the Younger Dryas observed in the Cariaco basin record is a faithful reflection of a $\Delta^{14}C$ change in the atmosphere at that time, the previous concern to explain the early $\Delta^{14}C$ drawdown during the Younger Dryas should be substituted by a new concern to explain this increase."*

Also, the size of the anomaly agrees generally with prediction: it is the largest anomaly, and the closest SN (Table 1). The timing, proximity, and energy of this prehistoric SN, when compared to the rapid increase, size, global extent, and timing of the $^{14}C$ anomaly, supports a possible cause and effect relationship (Brakenridge 2011). The initial $^{14}$C rise may, however, have been enhanced by climate-related carbon cycle and other Earth system-internal changes.

*G330.0+15.0, Lupus Loop, 22500 BP*. This is a SNR that may be nearly as close (.32 kpc) as Vela, but is older: 23000 ± 8000 BP (Table 1) (Safi-Harb et al. 2012). The time period experienced a steep 21‰ $\Delta^{14}$C increase over 140 y at 22500 BP (Figure 1b) but, unlike the case for Vela's approximate age, there is no close-interval $^{14}$C sampling available. Neither the distance or age of this SN are well constrained; a revised Σ–D relation, for example, estimates its distance at 0.5 kpc (Pavlovi ́c et al. 2014); an age of 50,000 BP may be consistent with the X-ray observations. Without more narrow SN age constraints and detailed $^{14}$C sampling, no confirmation or falsification of a correlated rapid-increase $^{14}$C signal from this nearby SN is possible. However, the Intcal13 data in figure 1b is compatible with a signal of the expected size and at an appropriate time.

*G114.3+00.3, S165, 7431 BP*. IntCal13 results showing a steep rise of 15.3‰ $\Delta^{14}$C between 7440 and 7410 BP are provided in Figure 1c, but detailed $^{14}$C assays for this period from Bristlecone Pine are also available with a 1-2 y resolution. They demonstrate a large increase (20‰) over ten years, from 7431 to 7421 BP (Miyake et al. 2017). Recently, the distance of S165 was revised downward to ~.7 kpc and with an age of approximately 7500 BP; the total energy is estimated at 5 x 10$^{51}$ ergs as a Type II event (Yar-Uyaniker et al. 2004). The distance is from associated patches of H I and H II emission (Safi-Harb et al. 2012), and there is also a central pulsar. The rapid rise, magnitude, and duration of the $^{14}$C anomaly is compatible with causation by this close and powerful SN source, but abnormal solar activity has instead been invoked (Jull et al. 2018; Miyake et al. 2017). However, S165 is of appropriate age and distance to have caused this anomaly. Thus, the estimated $^{14}$C production rate over one year is 1.2 – 11.2 a/cm$^2$/s (range due in part to the distance uncertainty); and box modeling and comparison to solar modulation effects of the $^{14}$C anomaly indicate a total $^{14}$C increase in production of between 6.0 ± 2.4 and 10.5 ± 3.0 a/cm$^2$/s (Miyake et al. 2017). Predicted SN effects therefore appear to be in agreement with the tree ring data.

*G266.2-1.2, Vela Jr., 2765 BP.* The IntCal13 radiocarbon chronology shows a $\Delta^{14}$C rise of 20‰ (figure 1d, 1e) at 2765 BP. This anomaly has recently been further investigated by more detailed tree ring $^{14}$C data demonstrating a very rapid rise of approximately 13‰ at 2765-2749 BP (Jull et al. 2018). Vela Jr is a shell-type SNR in the same line of sight as G263.9-03.3 Vela. Its age is now estimated at 3800 ± 1400 y and distance at .75 ± .25 kpc (Allen et al. 2015). No associated pulsar or other compact object has so far been observed. The age is between 2400 and 5100 y if it is expanding into a uniform ambient medium; if it is instead expanding into the material shed by a steady stellar wind, then the age may be as much as 50% older. SN causation for the brief, rapid-increase $^{14}$C anomaly at 2765 BP is supported by the presence of this object, which is of approximately the correct age and expected γ intensity (Table 1).

*G160.9+02.6, HB9, 5340 BP.* IntCal13 results indicate a +18‰ $^{14}$C anomaly commencing at 5340 BP (table 1 and figure 1d), but a rise that is not as steep as others, to ~5280 BP. However, a detailed tree ring study (Wang et al. 2017) concludes an abrupt rise (+9‰ in one year) in 5372 BP with a decay period of about 10 y; this result, from a floating tree ring chronology and buried logs, has not yet been validated at other sites. HB9 is a radio SNR with an associated magnetar/pulsar compact object. The age is estimated at 4000-7000 BP based on the Sedov equation and evaporative cloud modeling (Leahy and Tian 2007); it is nearly as close as Vela Jr. at .80 ± .40 kpc. HB9 is of appropriate distance and age to be compatible with this brief $^{14}$C anomaly.

*G106.3+02.7, Boomerang, 10255 BP.* At this time, a 12‰ rise in $\Delta^{14}$C occurs within 35 y in the IntCal13 results (figure 1f). The steep rise is followed by a more gradual but sustained rise to 10145 BP (another 10‰), possibly of different causation. The Boomerang SNR is ~10000 y in age and is at a distance of ~.8 kpc. Table 1 provides another SN, G89.0+4.7, with an appropriate age (9900 ± 5100 BP) but at greater distance (1.25 ± .45 kpc). The larger distance implies a smaller $^{14}$C effect. Boomerang is more consistent with the measured anomaly at 10255 BP (Table 1). Single year tree ring analysis is needed to further constrain the characteristics of $^{14}$C through this time interval.

*G107.5-1.5*, 4880 BP. The IntCal13 record includes a brief, positive (17.6‰ rise in $\Delta^{14}$C) anomaly at 4880-4820 BP (figure 1d), but the presence of a 1-2 y rapid-rise anomaly cannot be determined from these data alone. However, this relatively newly-discovered SN is close to the Earth, at 1.1 kpc, and may be of appropriate age: 4500 ± 1500 (Kothes 2003). As for Boomerang, single year tree ring analysis is needed to further constrain the characteristics of $^{14}$C through this time interval.

*G074.0-08.5, Cygnus Loop, 14722 BP.* The IntCal13 record bracketing this time reveals no brief anomalies (figure 1g), but floating tree ring records document a relatively brief episode of much-increased atmospheric $^{14}$C concentration (Adolphia et al. 2017) (compare figures 1g and 1h, which shows the change, per the source reference, as a period of much-younger $^{14}$C ages). The Hulu Cave speleothem data from China also support a brief (10 y) atmospheric $^{14}$C excursion (Southon et al. 2012). The $\Delta^{14}$C increase occurs at the beginning of another short-lived but geographically extensive cold interval in climate history: the Older Dryas Stadial (Mangerud et al. 2017). The Cygnus Loop exhibits a radio and X-ray shell (Fesen et al. 2018a) and its distance is now constrained to .74 ± .03 kpc (Fesen et al. 2018b). If the true SN age is close to 15,000 BP, then the Cygnus Loop SN is a candidate for causation of the brief 14722 BP $^{14}$C anomaly (Table 1).

The other SNe listed in Table 1 are at larger distances of approximately 1-1.4 kpc, and with the addition of SN 1006 at 1.6 kpc (for which a possible $^{14}$C signal has been recorded). These more-distant SNe also may be detectable by $^{14}$C assays with fine-scale temporal resolution and if the γ emitted was sufficiently energetic. In this regard, two other rapid-increase, global, and short-lived $^{14}$C anomalies have been identified at 1176-1175 and 957-956 BP in single-ring tree ring assays (Miyake et al. 2013; Miyake et al. 2012). The shapes of the $^{14}$C time series are similar: rapid increase within 1-2 y followed by a decade-long decay that could reflect operation of the carbon cycle. The magnitude of the younger event is 0.6 of the older, suggesting, if intrinsic energies of associated SNe were similar, that the younger SN was ~1.4 x more distant. Plausible candidate SNe would be *G347.3-00.5* at 1.3 kpc and the recently discovered and elongated *G190.9-2.2* at 1.0 ±0.3 kpc. This is closer than the SN 1006 SNR.

The *G190.9-2* remnant is of a similar mean radius, and thus age, as W49B: a SNR at 8 kpc distance that may record a GRB remnant (Lopez et al. 2013). This event is described by (Pavlov et al. 2013) as possibly producing the 1176 BP $^{14}$C anomaly. Using the 130 $^{14}$C a/erg production rate, a d = 1 km SN release of 4 x 10$^{49}$ erg of γ causes a one year addition of .9-2.4 a/cm$^2$/s (Table 1), whereas box modeling of the needed additional $^{14}$C needed indicates 3.9 a/cm$^2$/s (Miyake et al. 2013; Pavlov et al. 2013; Usoskin et al. 2013). If the event was as close at .75 kpc and emitted more energy (6 x 10$^{49}$), then the predicted $^{14}$C production matches that observed. This SN, like W49b, also lacks a pulsar central object, is elongated, and its shape suggests that the development of very energetic shock breakout γ accompanying a failed jet is possible. If a SNe causation is the case, it appears that this recently discovered SN may be stronger candidate than W49B. In any case, a possible historical sighting of this event also occurred in CE 774: a "red cross in the sky" in the Anglo-Saxon Chronicle (Allen 2012; Lovett 2012). This is compatible with the SN's location in the northern sky, and a non-point source optical object agrees with the observed complex-ringed appearance of the SN1987A remnant in very early stages of its evolution (Chevalier 1992).

## 6. Discussion and Conclusions

This paper demonstrates the viability of a SNe causation for many of the recorded rapid-increase and brief $^{14}$C anomalies in terrestrial records. All of the described anomalies may, alternatively, have a different causation, such as extreme solar flares (Miyake and al 2014), or, in the case of the Younger and Older Dryas changes, the effects of climate, biosphere, and ocean circulation effects on the global carbon cycle. However, these hypotheses encounter difficulties: the exceptionally intense solar flares needed have not been observed in historic times, and the suddenness of, for example, the 12760 BP $^{14}$C increase is difficult to model through Earth system-internal changes without increases in the atmospheric production rate. In contrast, SNe causation is supported by not only the predicted effects of galactic SNe in general, but also by the known occurrence of close objects of appropriate age.

Given the variety of SNe objects and their associated intrinsic luminosities and released γ radiation, their distances only partly determine the expected terrestrial $^{14}$C signals. In this regard, however, the predicted and measured sizes of the $^{14}$C anomalies generally agree also with a causal connection, because the closer events are associated with the larger signals. Thus: 1) The older Vela SNR (G263.9-03.3) is very much the closest to the Earth, its age is also relatively well-constrained, and the largest of the recorded rapid-increase $^{14}$C anomalies occurs at the appropriate time. 2) A set of 5 SNRs at .7 - .8 kpc distance each match in time smaller $^{14}$C anomalies observed in tree ring wood cellulose assays, on materials dated through dendrochronology sampling at high temporal resolution (Table 1 and above text). Note also that four other SNs at similar distances are much older, and not clearly associated with $^{14}$C anomalies (no tree-ring data of appropriately close-interval sampling are available): G040.5+00.5, G205.5+0.5, G180.0-1.7, and G119.5+10.2 in Table 1. These may also have left $^{14}$C signatures further back in time, but this must be determined by future work.

In conclusion, these data and analysis do not rule out solar flare or other causal hypotheses (Jull et al. 2018; Usoskin et al. 2013) for the rapid-onset $^{14}$C increases. However, SNe causation is compatible with present knowledge of the size of γ emissions which may be associated with the variety of SNe, and with the known distances and ages of a set of relatively close and young galactic SNRs. SN γ energies adequate to have produced the $^{14}$C pulses were much earlier predicted from theory; they have now been directly observed for SNe outside of our galaxy. If SNe causation of the cosmogenic isotope changes is actually not the case, then the detailed $^{14}$C record now emerging from (mainly) tree ring studies may provide a useful record of pre-historic solar variability and mega-flare production (Miyake et al. 2017); this record should be of societal concern. If SNe, instead, caused many or all of the rapid-increase $^{14}$C changes, then the radiation hazard remains, but is of a different nature.

## 7. Acknowledgement

## 8. References Cited

Adolphia F, Muscheler R, Friedrich M, Güttlere L, Talamo S, Kromerc B. 2017. Radiocarbon calibration uncertainties during the last deglaciation: Insights from new floating tree-ring chronologies. Quaternary Science Reviews. 170:98-108.

Allen GE, Chow K, DeLaney T, Filipović4 MD, Houck JC, Pannuti TG, Stage MD. 2015. On the expansion rate, age, and distance of the supernova remnant g266.2-1.2 (vela jr.). The Astronomical Journal. 798(2).

Allen J. 2012. Astronomy: Clue to an ancient cosmic-ray event? Nature. 486.

Brakenridge GR. 2011. Core-collapse supernovae and the younger dryas/terminal rancholabrean extinctions. Icarus. 215(1):101-106.

Cano Z. 2014. Gamma-ray burst supernovae as standardizable candles. The Astrophysical Journal. 794(121):9.

Cano Z, Wang S-Q, Dai Z-G, Wu X-F. 2017. The observer's guide to the gamma-ray burst supernova connection. Advances in Astronomy. 2017:41.

Cha AN, Sembach KR, Danks AC. 1999. The distance to the vela supernova remnant. The Astrophysical Journal Letters. 515(1):L25-L28.

Chevalier RA. 1992. Supernova 1987a at five years of age. Nature. 355:691-696.

Churazov E, Sunyaev R, Isern J, Bikmaev I, Bravo E, al e. 2015. Gamma rays from type ia supernova sn 2014. The Astrophysical Journal. 812(1).

Colgate SA. 1975. The prompt effects of supernovae. Ann NY Acad Sci. 262:34-46.

Damon PE, Kaimei D, Kocharov G, Mikheeva I, Peristykh A. 1995. Radiocarbon production by the gamma-ray component of supernova explosions Radiocarbon. 37(2):599-604.

Dodson R, Legge D, Reynolds JE, McCulloch PM. 2003. The vela pulsar's proper motion and parallax derived from vlbi observations. The Astrophysical Journal,. 596(2):1137-1141.

Fesen RA, Neustadt JMM, Black C, Milisavljevic D. 2018a. A distance estimate to the cygnus loop based on the distances to two stars located within the remnant. Monthly Notices of the Royal Astronomical Society,. 475(3):3996-4010.

Fesen RA, Weil KE, Cisneros IA, Blair WP, Raymond JC. 2018b. The cygnus loop's distance, properties, and environment driven morphology. Monthly Notices of the Royal Astronomical Society. 481:1786-1798.

Firestone RB. 2014. Observation of 23 supernovae that exploded <300 pc from earth during the past 300 kyr. The Astrophysical Journal. 789(29):11.

Fishman GJ, Inan US. 1988. Observation of an ionospheric disturbance caused by a gamma-ray burst. Nature. 331:418-420.

Gehrels N, Laird CM, Jackman CH, Cannizzo JK, Mattson BJ. 2003. Ozone depletion from nearby supernovae. Astrophys J. 585:1169-1176.

Gehrels N, Mészáros P. 2012. Gamma ray bursts. Science. 337:932.

Green DA. 2014. A catalogue of 294 galactic supernova remnants. Bull Astr Soc India. 42:47-58.

Hua Q, al e. 2009. Atmospheric 14c variations derived from tree rings during the early younger dryas. Quaternary Science Reviews. 28:2982-2990.

Hughen KA, Southon JR, Lehman SJ, Overpeck JT. 2000. Synchronous radiocarbon and climate shifts during the last deglaciation. Science. 290(5498):1951-1954.

Jiang S-Y, Zhao F-Y. 2007. The historical re-brightening and distance recheck of sn 1006. Chinese Journal of Astronomy and Astrophysics. 7(2).

Jull AJT, Panyushkina I, Miyake F, Masuda K. 2018. More rapid 14c excursions in the tree-ring record: A record of different kind of solar activity at about 800 bc? Radiocarbon.

Kann DA, Schady P, Olivares FE, Klose S, Rossi A, al e. 2018. Highly luminous supernovae associated with gamma-ray bursts, 1,grb 111209a/sn 2011kl in the context of stripped-envelope and superluminous supernovae. Astronomy & Astrophysics.

Kanu AM, Comfort LL, Guilderson TP, Cameron-Smith PJ, Bergmann DJ, Atlas EL, Schauffler S, Boering KA. 2016. Measurements and modeling of contemporary radiocarbon in the stratosphere. Geophysical Research Letters.

Kasen D, Woosley SE. 2009. Type ii supernovae: Model light curves and standard candle relationships. Astrophys J. 703(2):2205-2216.

Klein RI, Chevalier RA. 1978. X-ray bursts from type ii supernovae. The Astrophysical Journal. 223:L109-L112.

Kothes R. 2003. G107.5-1.5, a new snr discovered through its highly polarized radio emission. Astronomy and Astrophysics. 408(1).

Kovaltsov GA, Mishev A, Usoskin IG. 2013. A new model of cosmogenic production of radiocarbon 14c in the atmosphere. Earth and Planetary Science Letters. 337-338:114-120.

Kromer B, Friedrich M, Hughen KA, Kaiser F, Remmele S, Schaub M, Talamo S. 2004. Late glacial 14c ages from a floating, 1382-ring pine chronology. Radiocarbon. 46:1203-1209.

Leahy DA, Tian WW. 2007. Radio spectrum and distance of the snr hb9. Astron Astrophys. 461:1013-1018.

Lingenfelter RE, Ramaty R. 1970. Astrophysical and geophysical variation in c-14 production. In: Olsson IU, editor. Radiocarbon variations and absolute chronology. NewYork: John Wiley & Sons. p. 513-537.

Lopez LA, Ramirez-Ruiz E, Castro D, Pearson S. 2013. The galactic supernova remnant w49b likely originates from a jet-driven, core-collapse explosion. ApJ.

Lovett RA. 2012. Ancient text gives clue to mysterious radiation spike. Nature. 486.

Mangerud J, Briner JP, Goslar T, Svendsen JI. 2017. The bølling-age blomvåg beds, western norway: Implications for the older dryas glacial re-advance and the age of the deglaciation. Boreas. 46:162-184.

Matz SM, Share GH, Leising MD, Chupp EL, Vestrand WT, Purcell WR, Strickman MS, Reppin C. 1988. Gamma-ray line emission from sn1987a. Nature 331:416-418.
Menjo H, Miyahara H, Kuwana K, K. M, Muraki Y, Nakamura T. 2005. Possibility of the detection of past supernova explosion by radiocarbon measurement. In: Acharya BS, editor. Proceedings of the 29th international cosomic ray conference. Mumbai: Tata Institute of Fundamental Research. p. 357-360.
Miyake F, al e. 2014. Cosmic ray event of a.D. 774–775 shown in quasi-annual10be data from the antarctic dome fuji ice core. Geophysical Research Letters. 42:84-89.
Miyake F, Jull AJT, Panyushkinad IP, Wackere L, Salzerd M, Baisand CH. 2017. Large 14c excursion in 5480 bc indicates an abnormal sun in the mid-holocene. 114. 5:881-884.
Miyake F, Masuda K, Hakozaki M, Nakamura T, Tokanai F, Kato K, Kimura K, Mitsutani T. 2014. Verification of the cosmic-ray event in ad 993-994 by using a japanese hinoki tree. Radiocarbon. 56:1189-1194.
Miyake F, Masuda K, Nakamura T. 2013. Another rapid event in the carbon-14 content of tree rings. Nature Communications. 4:1748.
Miyake F, Masuda K, Nakamura T, Kimura K, Hakozaki M, Jull AJT, Lange TE, Cruz R, Panyushkina IP, Baisan C et al. 2016. Search for annual 14c excursions in the past. Radiocarbon.1-6.
Miyake F, Nagaya K, Masuda K, Nakamura T. 2012. A signature of cosmic-ray increase in ad 774–775 from tree rings in japan. Nature. 486:240-242.
Moriya TJ, Sorokina EI, Chevalier RA. 2018. Superluminous supernovae. Astrophys J.
Nakar E, Sari R. 2010. Early supernovae light curves following the shock breakout. The Astrophysical Journal. 725(1):904-921.
Nakar E, Sari R. 2012. Relativistic shock breakouts-a variety of gamma-ray flares: From low-luminosity gammay-ray bursts to type ia supernovae. The Astrophysical Journal. 747:15.
Olivier M, Stocker TF, Muscheler R. 2001. Atmospheric radiocarbon during the younger dryas: Production, ventilation, or both? Earth and Planetary Science Letters. 281:383-395.
Pavlov AK, Vdovina MA, Vasilyev GI, Pavlov AK, Blinov AV, Ostryakov VM, Konstantinov AN, Volkov PA. 2013. Ad 775 pulse of cosmogenic radionuclides production as imprint of a galactic gamma-ray burst. Monthly Notices of the Royal Astronomical Society.
Pavlović MC, Dobardžić A, Vukotić B, Urošević D. 2014. Updated radio $\sigma - d$ relation for galactic supernova remnants. Serb Astron J 189(25):1-5.
Pian E, Mazzali PA, Starling R. 2006. An optical supernova associated with the x-ray flash xrf 060218. Nature. 442:1011-1017.
Pinto PA, Woosley SE. 1988. The theory of gamma-ray emergence in supernova 1987a. Nature. 333:534-537.
Podsiadlowski P. 2013. Supernovae and gamma ray bursts. In: Oswalt T, editor. Planets, stars, and stellar systems. Springer. p. 693-733.
Reimer P. 2013. Intcal13 and marine13 radiocarbon age calibration curves 0–50,000 years cal bp. Radiocarbon. 55(4):1869-1887.
Ruderman MA. 1974. Possible consequences of nearby supernova explosions for atmospheric ozone and terrestrial life. Science. 184:1079-1081.
Safi-Harb S, Ferrand G, Matheson H. 2012. A high-energy catalogue of galactic supernova remnants and pulsar wind nebulae. In: Leeuwen Jv, editor. Neutron stars and pulsars: Challenges and opportunities after 80 years. Proceedings, IAU Symposium No. 291.

Scalo J, Wheeler JC. 2002. Astrophysical and astrobiological implications of gamma-ray burst properties. AstrophysJ. 566:723-737.
Southon J, Noronha AL, Cheng H, Edwards RL, Wang Y. 2012. A high-resolution record of atmospheric 14c based on hulu cave speleothem h82. Quaternary Science Reviews. 33:32-41.
Stuiver M, Polach HA. 1977. Discussion: Reporting of 14c data. Radiocarbon. 19(3):355-363.
Sushch I, Hnatyk B. 2014. Modelling of the radio emission from the vela supernova remnant. Astronomy and Astrophysics. 561(A139):8.
Thomas BC, Melott AL, Jackman CH, Laird CM, Medvedev MV, Stolarski RS, Gehrels N, Cannizzo JK, Hogan DP, Ejzak LM. 2005. Gamma-ray bursts and the earth: Exploration of atmospheric, biological, climatic, and biogeochemical effects. The Astrophysical Journal.
Usoskin IG, Kromer B, Ludlow F, Beer J, Friedrich M, Kovaltsov GA, Solanki SK, Wacker L. 2013. The ad775 cosmic event revisited: The sun is to blame. Astron Astrophys Lett. 552(L3).
Wang FY, Yu H, Zou YC, Dai ZG, Cheng KS. 2017. A rapid cosmic-ray increase in bc 3372–3371 from ancient buried tree rings in china. Nature Communications. 8:1487.
Yar-Uyaniker A, Uyaniker B, Kothes R. 2004. Distance of three supernova remnants from h i line observations in a complex region: G114.3+0.3, g116.5+1.1, and ctb 1 (g116.9+0.2). The Astrophysical Journal. 616(1):247-256.
A. Ray & R. A. McCray e, editor. Distances of galactic supernova remnants. Supernova Environmental Impacts; 2013.

**Table 1**. Distances, Earth-incident γ (using 4 x $10^{49}$ ergs total γ SN emission), ages, predicted $^{14}$C production, and measured $^{14}$C rise (reported as +Δ $^{14}$C) within the time intervals for 18 of the closest SNe. $^{14}$C production is based on 130 atoms per SN-generated γ erg. *Type Ia SN; all others are core collapse SNe. The energies and production ranges are based on the distance uncertainties. The Δ $^{14}$C results use the higher temporal resolution data when available, and as described in text. Ages are years before present ("BP", before CE 1950). All errors are expressed as standard errors.

**Figure 1a-h. Radiocarbon variation at the times of nearby prehistoric SNe. a.** A steep 25.3‰ rise in Δ$^{14}$C occurs in IntCal13 at 12745-12640 BP. **b.** A 21‰ IntCal13 Δ$^{14}$C rises occurs at 22500 BP, but the coarse temporal resolution is inadequate to reveal brief anomalies. **c.** IntCal13 shows a steep rise in Δ$^{14}$C of 15.3‰ at 7440-7410 BP. Single year tree ring data show a 20‰ rise from 7431 to 7421(Miyake et al. 2017). **d**. At least three rapid-increase $^{14}$C anomalies may be illustrated in this IntCal13 plot spanning 5500-2500 BP: one at 5340 BP, one at 4880 BP, and one at 2765 BP. **e.** At 2765-2735 BP, Δ$^{14}$C rises by 11.4‰ in 30 y, at the approximate time of the Vela Jr. SN.

**f.** Between 10255 and 10220 BP $\Delta^{14}$C rises by 12.2‰ at the approximate time of the Boomerang SN. **g.** IntCal13 data for 15000-13000 are without detailed time resolution and can reveal no short-lived anomalies. **h.** Floating tree-ring chronologies with closer temporal sampling, however, document a strong and short-lived $\Delta^{14}$C increase marked by younger radiocarbon dates just after 14722 BP (Adolphia et al. 2017) and compatible with the age and probable γ intensity of the Cygnus Loop SNR.